\documentclass[twoside,fleqn]{article}
\usepackage{espcrc2}
\usepackage{epsfig}
\usepackage{psfig}
\newcommand{\beqn} {\begin{equation}}
\newcommand{\eqn} {\end{equation}}

\newcommand{\plaq}{\mbox{\raisebox{-.75mm}
{\epsfig{file=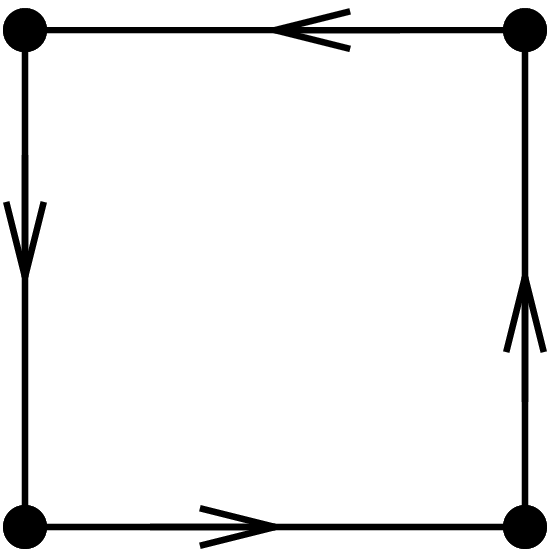,height=3mm
}}~}}

\newcommand{\tr}{\mbox{Tr~}}
\newcommand{\re}{\mbox{Re~}}
\newcommand{\kronecker}[2]{  \delta_{#1 , #2}  }

\newcommand{\mymatrix}[2]{
        m \kronecker{#1}{#2} + \left( \;
                    \frac{9}{16} A\left[ U \right]_{#1\; #2} -
        \frac{1}{48} B\left[ U  \right]_{#1\; #2} \right) }
\newcommand{\linkdagger}[2]{  U_{ #1 , #2}^\dagger  }
\newcommand{\link}[2]{  U_{ #1 , #2}  }
\newcommand{\apart}[3]{
\sum_{#3} \eta_{{#1},{#3}}\left( \link{#1}{#3} \kronecker{#1}{#2-\hat{#3}}
                -\linkdagger{#1-\hat{#3}}{#3}
                                        \kronecker{#1}{#2+\hat{#3}} \right) }

\newcommand{\bparta}[3]{
\sum_{#3} \eta_{{#1},{#3}}( \link{#1}{#3} \link{#1+\hat{#3}}{#3} 
                                       \link{#1+2 \hat{#3}}{#3}
                                             \kronecker{#1}{#2- 3 \hat{#3}}   }
\newcommand{\bpartb}[3]{
                 \linkdagger{#1-\hat{#3}}{#3} \linkdagger{#1-2\hat{#3}}{#3}
                 \linkdagger{#1-3 \hat{#3}}{#3} \kronecker{#1}{#2+ 3 \hat{#3}}
                  ) }

\newcommand{\AmS}{{\protect\the\textfont2
  A\kern-.1667em\lower.5ex\hbox{M}\kern-.125emS}}

\title{
\vskip -100pt
\mbox{} \hfill BI-TP 97/19\\
\mbox{} \hfill June 1997\\
\vskip 45pt
On QCD Thermodynamics with Improved Actions 
\thanks{Talk given at the International Workshop ``Lattice QCD on
Parallel Computers'', Tsukuba, Japan, 10-15 March 1997; to be published in
Nucl. Phys. B Supplements}
}
\author{Frithjof Karsch 
\\
\vskip 6pt
Fakult\"at f\"ur Physik, Universit\"at Bielefeld,
D-33615 Bielefeld, Germany}
\begin{document}
\begin{abstract}
%
We discuss recent
advances in the calculation of thermodynamic observables using improved
actions. In particular, we discuss the calculation of the equation
of state of the $SU(3)$ gauge theory, the critical temperature in units
of the string
tension, the surface tension and the latent heat at the deconfinement
transition.  We also present first results from a calculation of
the equation of state for four-flavour QCD using an ${\cal O} (a^2)$
improved staggered fermion action and discuss possible further improvements
of the staggered fermion action.
\end{abstract}

\maketitle
\vskip 20pt
%
\noindent
\section{Introduction}

The idea that strongly interacting hadronic matter undergoes a phase
transition to a new phase, the quark-gluon plasma, has been around
for a long time. More than 10 years ago lattice calculations have
given first direct evidence for the existence of such a phase
transition and opened the way for its detailed quantitative analysis.
Many of the non-perturbative features of finite temperature QCD, which
originally emerged from perturbative studies 
have since then been analyzed in lattice calculations.
However, many of these numerical studies also had to remain on a qualitative
level for a long time because simulations on coarse lattices were
hampered by large discretization errors and calculations at smaller
lattice spacing suffered from poor statistical accuracy.

The discretization scheme proposed originally by K. Wilson for gauge
theories \cite{Wil74} does for a finite lattice cut-off, $a$, introduce 
systematic ${\cal O} (a^2)$ deviations from the continuum formulation.
It is well known that this introduces
severe problems in thermodynamic calculations, which in the past made a
direct determination of the physics in the continuum limit difficult. 
Already the thermodynamics of free Bose (gluon) or Fermi (quark) 
gases deviates strongly from the continuum ideal gas results when
calculated on coarse Euclidean lattices \cite{Eng82}. This problem carries 
over to QCD, where bulk thermodynamic observables like energy
density and pressure do rapidly come close to the
non-interacting ideal gas limit above the deconfinement phase transition. 
The calculation of these quantities is thus expected to suffer from
similarly strong cut-off effects as the ideal gas, despite the fact
that the high temperature plasma phase does in many respects still 
show strong non-perturbative properties, characterized e.g. by thermal 
screening masses and quasi-particle excitations.

At high temperature
the relevant contributions to thermodynamic observables result from
momenta which are of the order of the temperature, {\it i.e.} 
$\langle p \rangle \sim 3T$. However,
in lattice calculations $T$ is fixed through the
temporal extent, $N_\tau$, of the lattice  and the cut-off, $a$, {\it
i.e.} $T\equiv 1/N_\tau a$. As the computational effort
in the calculation of bulk thermodynamic observables, which are of dimension
$T^4$, increases approximately like $N_\tau^{10}$, it is evident that
the temporal extent of the lattice has 
to remain rather small in most thermodynamic calculations.
The relevant momenta for e.g. lattices with temporal extent $N_\tau=4$ 
are therefore close to the cut-off where the 
lattice and continuum dispersion relations differ strongly from each other. 
Indeed this is the origin of
the well-known discrepancy between the energy density of an ideal gas calculated
on a finite lattice ($\epsilon (N_\tau)$) and in the continuum 
($\epsilon_{\rm SB}$) \cite{Eng82}. 
The cut-off dependence in finite temperature calculations thus
shows up as a {\it finite size effect}, which should not be confused
with the finite size dependence resulting from the spatial extent of the
lattice. The latter is an infra-red effect and controls the approach to
the thermodynamic limit.

In the next section we briefly discuss some improved actions,
which recently have been used in thermodynamic calculations. In Section 3
we present results from a perturbative calculation of cut-off effects in the 
ideal gas (infinite temperature) limit. Results from numerical simulations of 
the $SU(3)$ gauge theory and four-flavour QCD are discussed in Sections 4 and 5. 
In Section 6 we suggest a new staggered fermion action suitable
for thermodynamic calculations. Finally we give our conclusions in 
Section 7.

\section{Improved actions}

During the last few years much progress has been made in dealing with the
systematic discretization errors in lattice regularized quantum field
theories. Various improved discretization schemes
for the Lagrangian of QCD have been constructed and explored, which do show
much less cut-off dependence than the prescription originally given by
Wilson \cite{Nie96,Nie97,Lep97}.

When formulating a discretized version of QCD one has a great deal of
freedom in choosing a lattice action. Different formulations may differ by
sub-leading powers of the lattice cut-off, which vanish in the continuum
limit. This has, for instance, been used by Symanzik to systematically
improve scalar field theories \cite{Sym83} and has then been applied 
to lattice regularized $SU(N)$ gauge theories \cite{Wei83,Lue85}. In
addition to the elementary
plaquette term appearing in the standard Wilson formulation of lattice QCD
larger loops can be added to the action in such a way that the leading
${\cal O} (a^2g^0)$ deviations from the continuum formulation are 
eliminated and corrections only start in ${\cal O} (a^4g^0, a^2g^2)$. 
A simple class of improved actions is, for instance, obtained by adding
planar loops of size $(k,l)$ to the standard Wilson action (one-plaquette 
action)

\begin{eqnarray}
S^{(k,l)}\hskip -0.3cm &=& \hskip -0.3cm
\sum_{x, \nu > \mu} \hskip -0.15cm \left(c_0^{(k,l)}~W_{\mu,\nu}^{1,1} (x) +
c_1^{(k,l)} W_{\mu,\nu}^{k,l} (x) \right)~,
\label{actions}
\end{eqnarray}
where $W_{\mu,\nu}^{k,l}$ denotes the average of planar Wilson loops of 
size $(k,l)$ and $(l,k)$ in the $(\mu,\nu)$ plane of a 4-dimensional lattice
of size $N_\sigma^3\times N_\tau$\cite{Bei96}. 
The coefficients  $c_i$ are, in general, functions of the gauge coupling
$g^2$, {\sl i.e.} $c_i \equiv c_i(g^2)$. In the limit $g^2 \rightarrow 0$
the two coefficients in Eq.~\ref{actions} should fulfill the relation
$c_0^{(k,l)}(0) = 1- k^2 l^2 c_1^{(k,l)} (0)$ in order to insure the 
correct continuum limit.
A specific choice of $c_1^{(k,l)}(0)$ allows to eliminate the leading 
${\cal O} (a^2)$ corrections on the 

\vspace{0.2cm}

\noindent
tree level: 
$\displaystyle{c_1^{(k,l)} (0) = -{2 / \bigl[ k^2 l^2 (k^2 + l^2 -2)}\bigr] }$ 

\vspace{0.2cm}
In the following we will present results for the
tree level improved actions $S^{(1,2)}$ and $S^{(2,2)}$ which
have been used recently for thermodynamic calculations \cite{Bei96,Bei96a,Bei96b}. 
These actions
can be further improved by adding either additional loops, which    
would allow a systematic perturbative improvement in higher orders of $g^2$, 
or they may be improved non-perturbatively by decorating the expansion
coefficients with tadpole improvement factors \cite{Lep93} 

\vspace{0.2cm}

\noindent
$\displaystyle{c_1^{(k,l)} (0) \rightarrow c_{1,tad}^{(k,l)}
= u_0^{-2(k+l-2)} c_1^{(k,l)} (0)}$

\vspace{0.2cm}
Here the factor $u_0$ is  determined self-consistently, for instance 
from plaquette expectation values \cite{Lep93},
$
u_0^4 = {1 \over 6 N_\sigma^3 N_\tau}~
\langle \sum_{x, \nu > \mu}
(1-\frac{1}{N}\re\tr\plaq_{\mu\nu}(x)) ~\rangle.
$
Studies at $T=0$ have shown that this approach indeed leads to a strong
reduction of the cut-off dependence even on rather coarse lattices. This
is, of course, most efficient for observables which are sensitive to
short distance scales.  We will discuss here in how far
this improvement scheme helps to improve finite temperature observables. 

Another way of generating an improved action within the class of actions 
defined by Eq.~\ref{actions} has been suggested by Y. Iwasaki \cite{Iwa85}
and has been extensively used recently for thermodynamic studies 
\cite{Iwa96}. Here renormalization group (RG) methods have been
used to generate an action close to the renormalized trajectory, which 
then has been projected onto the two parameter space defined by 
Eq.~\ref{actions}. This action is defined by the choice 
\beqn
c_1^{(1,2)} \equiv c_{1}^{\rm RG} = -0.662~~.
\label{c1mcrg}
\eqn

Finally we should mention the class of fixed point actions, which has been
constructed recently \cite{Nie96} and which also have been used for 
thermodynamic calculations \cite{Pap96,DeG96}. The fixed point action 
reproduces the continuum dispersion relation and therefore does, by 
construction, also eliminate all cut-off dependences. These do return,
when the fixed point action, which depends on an infinite number of 
parameters, is projected on to a limited parameter space. A frequently used
truncated version is expressed in terms of powers of two basic Wilson
loops, the plaquette, $W_{\mu,\nu}^{(1,1)}$, and a non-planar six-link
Wilson loop, which we denote here by $W_{\mu,\nu,\rho}$. The action
then reads
\begin{eqnarray}
S^{\rm FP}\hskip -0.3cm &=& \hskip -0.3cm
\sum_{x, \nu > \mu} \sum_{n=1}^k d_{0,n}~
\bigl[W_{\mu,\nu}^{1,1} (x)\bigr]^n \nonumber \\
& & +
\sum_{x, \nu > \mu > \rho} \sum_{n=1}^k d_{1,n}~
\bigl[ W_{\mu,\nu,\rho}(x) \bigr]^n ~,
\label{actionsFP}
\end{eqnarray}
here $W_{\mu,\nu,\rho}(x)$ denotes the average over the four
distinct paths of length six on the boundary of a 3-d cube.
In the high temperature, ideal gas limit these truncated fixed point actions
are indeed very similar to the ones defined by Eq.~\ref{actions}. The
ideal gas limit receives contributions only from the $n=1$ terms in the
above sums, {\it i.e.} it is controlled by the two couplings $d_{0,1}$
and $d_{1,1}$ only. The ansatz given by Eq.~\ref{actionsFP} may also be 
used to define a tree level improved action by choosing $k=1$ and 
$d_{0,1} = 1/3,~d_{1,1}= 1/3$. This is similar to Eq.~\ref{actions} with
the planar six-link loop replaced by a {\it twisted} six-link loop. 

\section{The High Temperature Ideal Gas Limit}

\begin{figure}[htb]
\begin{minipage}[t]{80mm}
\vspace{9pt}
   \epsfig{
       file=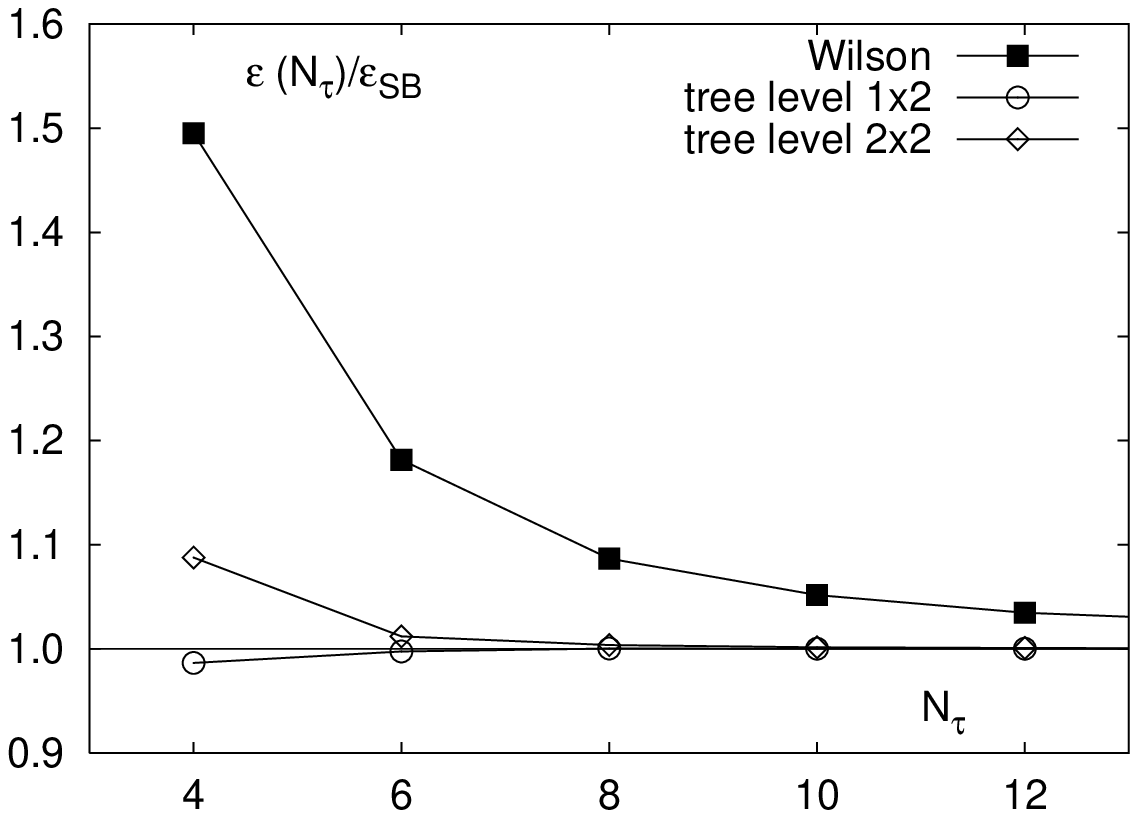,width=80mm}
\end{minipage}
%
%
\hskip -5pt
\begin{minipage}[t]{75mm}
\vskip 5pt
  \epsfig{
       file=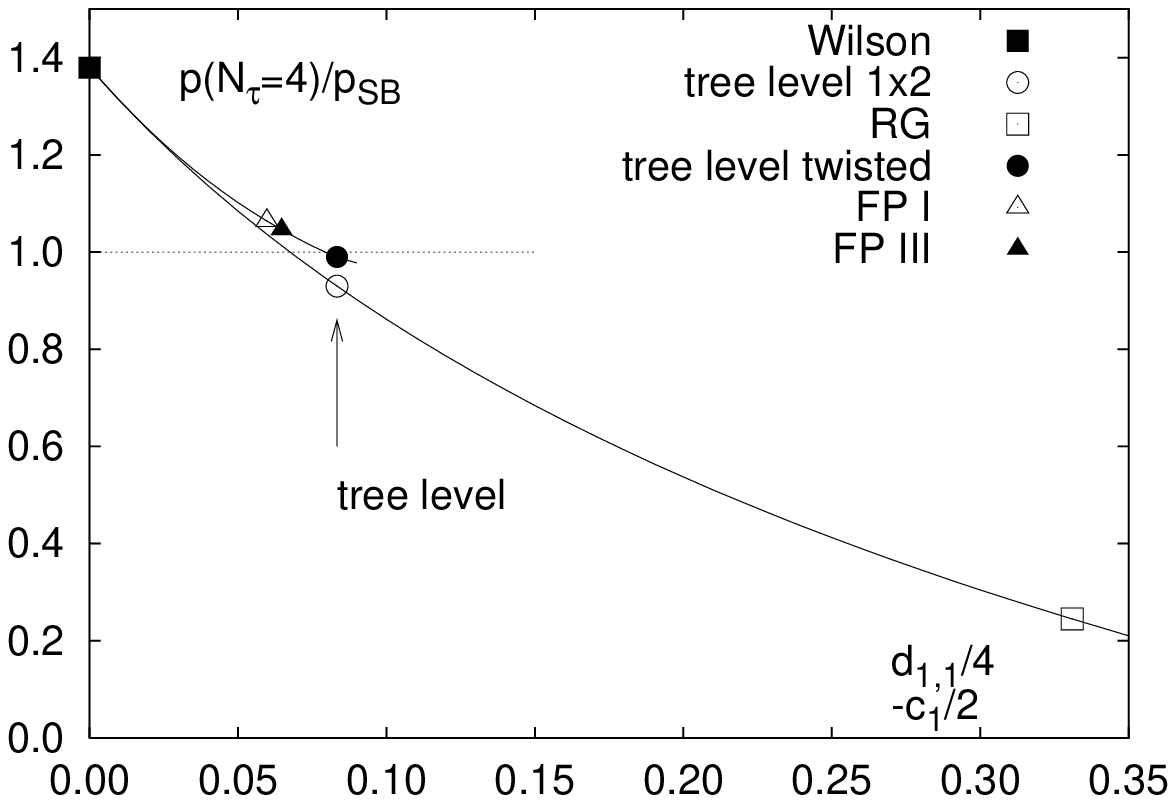,width=80mm}
\end{minipage}
\caption{Cut-off dependence of the energy density and pressure in the
infinite temperature, ideal gas limit. The upper figure shows results
for some actions as function of the temporal lattice extent $N_\tau$.
In the lower figure results are shown for $N_\tau=4$. For further discussions
see the text.} 
\label{fig:ideal}
\end{figure}

The importance of improved actions for thermodynamic calculations 
becomes evident immediately from an analysis of the high (infinite) 
temperature limit for a gluon gas on lattices with temporal extent $N_\tau$. 
In the standard Wilson formulation the ${\cal O}(a^2)$ cut-off dependence
in physical observables is reflected in large ${\cal O}(N_\tau^{-2})$ 
deviations from the continuum Stefan-Boltzmann limit. 
On lattices with temporal extent $N_\tau$ one finds, for instance, 
for the deviation of the energy density, $\epsilon (N_\tau)$, 
from the continuum result, $\epsilon_{\rm SB}$, \cite{Bei96}\footnote{The
${\cal O}(N_\tau^{-4})$ coefficient has been given incorrectly in \cite{Bei96}. 
It should read 2/5 rather than 2/6. We thank T. Scheideler for pointing this
out to us.}

\beqn
{\epsilon (N_\tau) \over \epsilon_{\rm SB} } = 1+
{10  \over 21}  \bigl({ \pi \over N_\tau}\bigr)^2 +
{2 \over 5} \bigl({ \pi \over N_\tau}\bigr)^4 +
{\cal O} \bigl( N_\tau^{-6} \bigr) 
\label{wilsongas}
\eqn
The large ${\cal O}(a^2)$ corrections are eliminated by using the improved
actions defined in Eq.~\ref{actions}.
One indeed finds a strong reduction of the cut-off dependence relative 
to the standard Wilson formulation. In fact,
although one has no control over the sub-leading corrections when setting
up the systematic ${\cal O}(a^2)$ improvement it turns out that in many
cases not only the ${\cal O}(a^2)$ corrections get eliminated but also the
${\cal O}(a^4)$ corrections get reduced. 
For instance, one finds \cite{Bei96}
\beqn
{\epsilon (N_\tau) \over \epsilon_{\rm SB} } = 1+
c_I \cdot \bigl({ \pi \over N_\tau}\bigr)^4 +
{\cal O} \bigl( N_\tau^{-6} \bigr) 
\label{improvedgas}
\eqn
with $c_I= 0.044 (2)$ for the action $S^{(1,2)}$ and $c_I= 0.178 (2)$ 
for $S^{(2,2)}$.
This leads to a drastic reduction of cut-off errors as can be
seen from Fig.~\ref{fig:ideal}. Of course, in this infinite
temperature limit
tadpole improvement does not affect the thermodynamics. In how far it
leads to an improvement at finite temperatures will be discussed in the 
next section. 

In the general case of actions, which also include non-planar loops,
it is more suitable to calculate the pressure, $p$, (or free energy 
density, $f=-p$) rather than the energy density. The former is proportional 
to the logarithm of the partition function, and thus is easily 
obtained in perturbative calculations.
In the lower part of Fig.~\ref{fig:ideal} we also show the 
ratio ${p (N_\tau) / p_{\rm SB} }$ for $N_\tau=4$ for several actions defined 
through Eqs.~\ref{actions} and \ref{actionsFP} with varying 
coefficients $c_1$ and $d_{1,1}$, respectively. 
This also includes the case of the
RG-improved action defined through Eq.~\ref{c1mcrg} (open square). As can be 
seen this action leads to rather large cut-off dependent corrections in the  
high temperature limit. In this figure we also show results for two versions
of the fixed point action\footnote{We use the notation FP I and FP III for the
two sets of fixed point actions defined in \cite{Pap96}.} \cite{Pap96} 
(triangles) and the tree level improved
actions including planar (open circle) and twisted (filled circle) six-link
loops. As can be seen all these actions do lead to small deviations from
the continuum result already on lattices with temporal extent $N_\tau=4$.   

A similar systematic improvement as discussed above for the gluonic sector
can be achieved in the fermion sector \cite{Eng96}. We will present some 
results for four-flavour QCD in Section 5.
In the next section we start with a discussion of results obtained for the
thermodynamics of the pure SU(3) gauge theory using mainly tree level and
tadpole improved actions \cite{Bei96,Bei96a,Bei96b}.

\section{SU(3) Thermodynamics}

Only recently calculations with the standard Wilson action could be
extended to lattices with sufficiently large temporal extent
($N_\tau =6$ and 8) that would allow
an extrapolation of lattice results for bulk
thermodynamic quantities to the continuum limit \cite{Eng95,Boy96}.
Computationally the step from $N_\tau =4$ to $N_\tau=8$ is quite non
trivial as the computer time needed to achieve numerical results with
the same statistical significance on a two times larger lattice
increases roughly like $2^{10}$. It therefore is highly desirable to use
improved actions, which suffer less from discretization errors, also for
thermodynamic calculations.

We have analyzed the thermodynamics of the SU(3) gauge theory
using the ${\cal O}(a^2)$ tree level ($u_0 \equiv 1$) and
tadpole
($u_0 < 1$) improved actions defined in Eq.~\ref{actions}.
As discussed above it seems to be plausible that improved actions will help
to reduce the cut-off dependence of thermodynamic quantities in the high
temperature  plasma phase. It is, however, less evident that this also is of
advantage for calculations close to the deconfinement phase transition where
the physics is strongly dependent on contributions from infrared modes. In
addition to a calculation of the equation of state an analysis of the
cut-off dependence of the critical temperature itself as well as
properties of the first order deconfinement transition like the surface
tension and latent heat
are therefore of interest.

\subsection{The critical temperature}

\begin{table*}[hbt]
\setlength{\tabcolsep}{1.5pc}
\catcode`?=\active \def?{\kern\digitwidth}

\caption{Critical temperature in units of $\sqrt{\sigma}$
on lattices with temporal extent $N_\tau=4$, {\it i.e.} at a value of the
cut-off given by $aT_c =0.25$. Infinite volume extrapolations for the 
critical couplings have been performed in all cases. Further details on the
data can be found in the references given in Figure 2.}
\vskip 5pt
\label{tab:tc}
\begin{center}
\begin{tabular}{|l|l|l|}
\hline
action &
$\beta_c$&
$T_c/\sqrt{\sigma} $\\ \hline
standard Wilson  &5.69254~(24)&$ 0.5983~(30)$\\
(2,2) (tree level improved) &4.3999~(3)&$ 0.625~(4) $\\
(1,2) (tree level improved) &4.0730~(3)&$ 0.633~(3) $\\
(1,2) (tadpole improved) &4.3525~(5)&$ 0.634~(3) $\\
(1,2) (RG improved) &2.2879~(11)&$ 0.653~(6)(1) $\\ \hline
\end{tabular}
\end{center}
\end{table*}

\begin{figure}[htb]
\vspace{-10pt}
   \epsfig{
       file=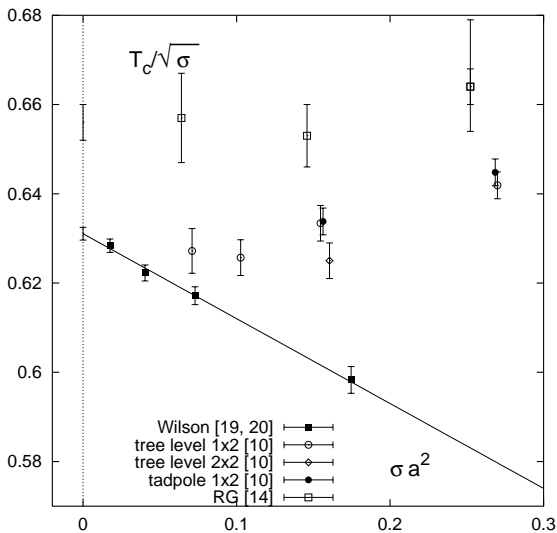,width=105mm}
\vskip -0.7truecm
\caption{The critical temperature in units of the square root of the string 
tension  for various actions versus the square of the cut-off. For the 
tree level improved $(1,2)$-action results for the critical coupling on
finite spatial lattices with temporal extent $N_\tau=5$ and 6 have been taken 
from Ref~\cite{Cel94} and extrapolated to infinite volume \cite{Bei96a}.}
\label{fig:string}
\end{figure}

When comparing the cut-off dependence of thermodynamic observables
one, of course, has
to make sure that these are compared at the same value of the cut-off.
On lattices with fixed temporal extent this amounts to a comparison at
the same value of the
temperature, which, for instance, can be defined in terms of  the critical
temperature for the deconfinement transition, {\it i.e.} $T/T_c$.
Also the determination of this
temperature scale is influenced by the finite lattice spacing and will
contribute to the overall cut-off dependence of thermodynamic observables.
An indication for the size of cut-off dependence in the definition of
a temperature scale can be deduced from a calculation of the critical
temperature in units of $\sqrt{\sigma}$. The ratio $T_c/\sqrt{\sigma}$ has
been studied in quite some detail at different values of the cut-off for 
the $SU(3)$ gauge theory with the Wilson action. The ${\cal O} (a^2)$ 
dependence has clearly been seen in these calculations.
An extrapolation of
these results to the continuum limit yields\footnote{This differs slightly
from the value published in \cite{Boy96}. The difference results from
a re-analysis of the string tension \cite{lego} at $\beta_c(N_\tau)$ for 
$N_\tau = 8$ and 12 which now is 
based on an interpolation of results obtained at nearby values of the 
gauge coupling rather than a single calculation performed at $\beta_c(N_\tau)$
as it has been done in \cite{Boy96}.}
\beqn
{T_c \over \sqrt{\sigma} } = 0.631 \pm 0.002~~.
\label{tcsigma}
\eqn
The reduction of the cut-off dependence may best be discussed by comparing 
calculations of $T_c/\sqrt{\sigma}$ at a fixed value of
the cut-off, for instance for $aT_c = 1/4$,  with the above continuum 
extrapolation. Such a comparison is performed in Table~\ref{tab:tc} for 
some tree level and tadpole improved actions as well as the RG-improved
action. While in the case of the Wilson action $T_c/\sqrt{\sigma}$ deviates 
by roughly 10\% from the 
continuum extrapolation at this value of the cut-off, the tree level and
tadpole improved actions only show little deviations. We also note that
the result for the tadpole improved action is consistent within errors 
with the tree level result. The RG-improved action does lead to a somewhat
larger value for $T_c/\sqrt{\sigma}$ than all the other calculations
presented in Table~\ref{tab:tc}. However, when comparing calculations at
different values of the cut-off (Fig.~\ref{fig:string}) 
we note that the RG-improved action as well as the other improved actions
shows no significant cut-off dependence, unlike the Wilson action where
the ${\cal O} (a^2)$ cut-off dependence is clearly visible.   
It thus seems that the discrepancy of about 4\% between the different improved 
action calculations is mainly due to differences in the determination of
the string tension from the heavy quark potential and not due to different
${\cal O} (a^2)$ corrections, {\it i.e.} the accuracy is at present limited
through ambiguities in extracting the long-distance physics from the 
heavy quark potential. This ambiguity is reflected in the spread of  
current determinations of the critical temperature,  
\vskip 5pt
$\displaystyle{
T_c/\sqrt{\sigma}}$ = 0.631~(2) \cite{Boy96,lego} --  0.656~(4) \cite{Iwa96}. 

\subsection{Bulk thermodynamics}

Bulk thermodynamic quantities like the energy density ($\epsilon$) or
pressure ($p$) calculated with the Wilson action on lattices of size
$N_\sigma^3  N_\tau$ with $N_\tau = 4$, 6 and 8 \cite{Boy96}
show a strong cut-off dependence in the plasma phase.
A first analysis with the tree level improved (2,2)-action \cite{Bei96}
has shown that this cut-off dependence gets drastically reduced already on 
an $N_\tau=4$ lattice. This is true also for the tree level improved 
(1,2)-action, the
tadpole improved (1,2)-action \cite{Bei96b} and a fixed point action
\cite{Pap96} which all
yield results consistent with the continuum extrapolation performed
for the standard Wilson action. As an example we discuss here
the pressure.

The pressure can be obtained from an integration of the difference of
action densities at zero
($S_0$) and finite ($S_T$) temperature \cite{Eng90},
\beqn
{p\over T^4}\Big\vert_{\beta_0}^{\beta}
=~N_\tau^4\int_{\beta_0}^{\beta}
{\rm d}\beta'  (S_0-S_T) ~,
\label{preslat}
\eqn
where the zero temperature calculations are performed on a large lattice of
size $N_\sigma^4$ and the finite temperature calculations are performed on
lattices of size $N_\sigma^3 N_\tau$. In order to compare calculations
performed with different actions one has to determine a physical temperature
scale. This can, for instance, be achieved through a calculation of
the string tension at several values of the gauge coupling ($\beta$) and at the
critical coupling for the phase transition on a lattice of temporal size $N_\tau$.
This yields $T/T_c \equiv \sqrt{\sigma}(\beta_c) / \sqrt{\sigma}(\beta)$.

\begin{figure}[htb]
\vspace{9pt}
\begin{minipage}[t]{80mm}
   \epsfig{
       file=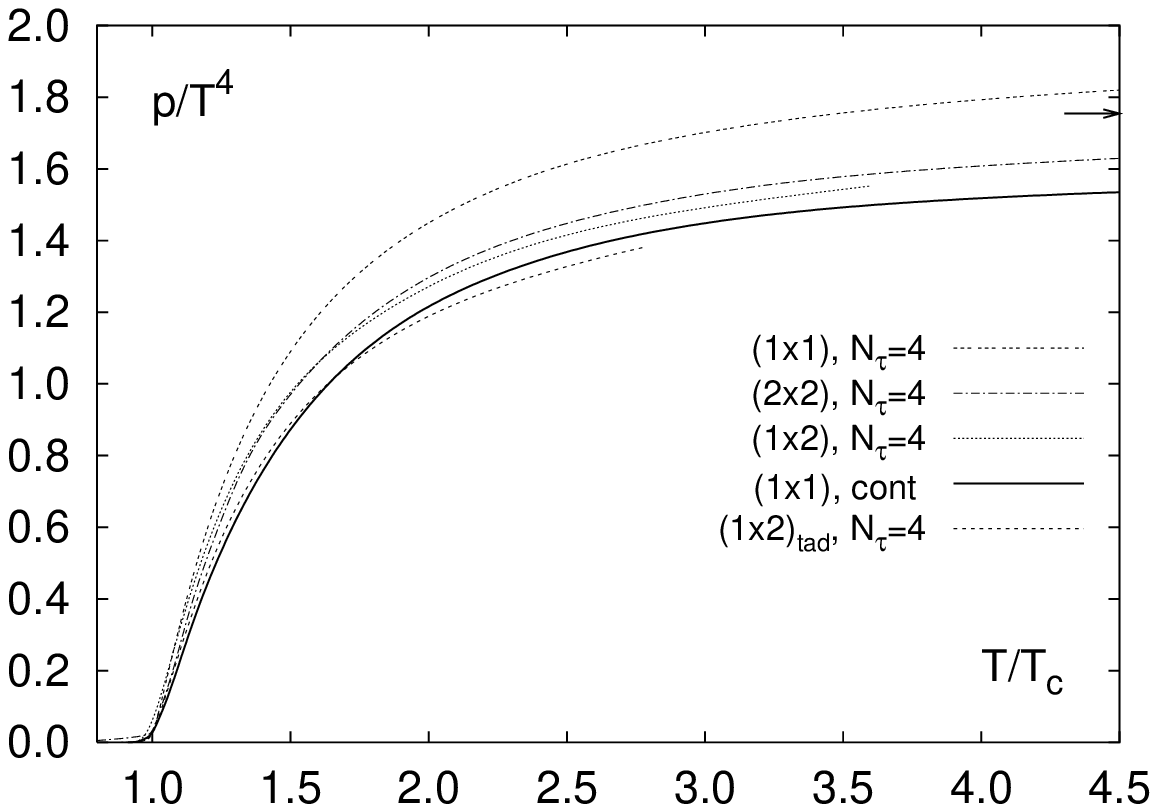,width=80mm}
\end{minipage}
%
%
\hskip -5pt
\begin{minipage}[t]{70mm}
  \epsfig{
       file=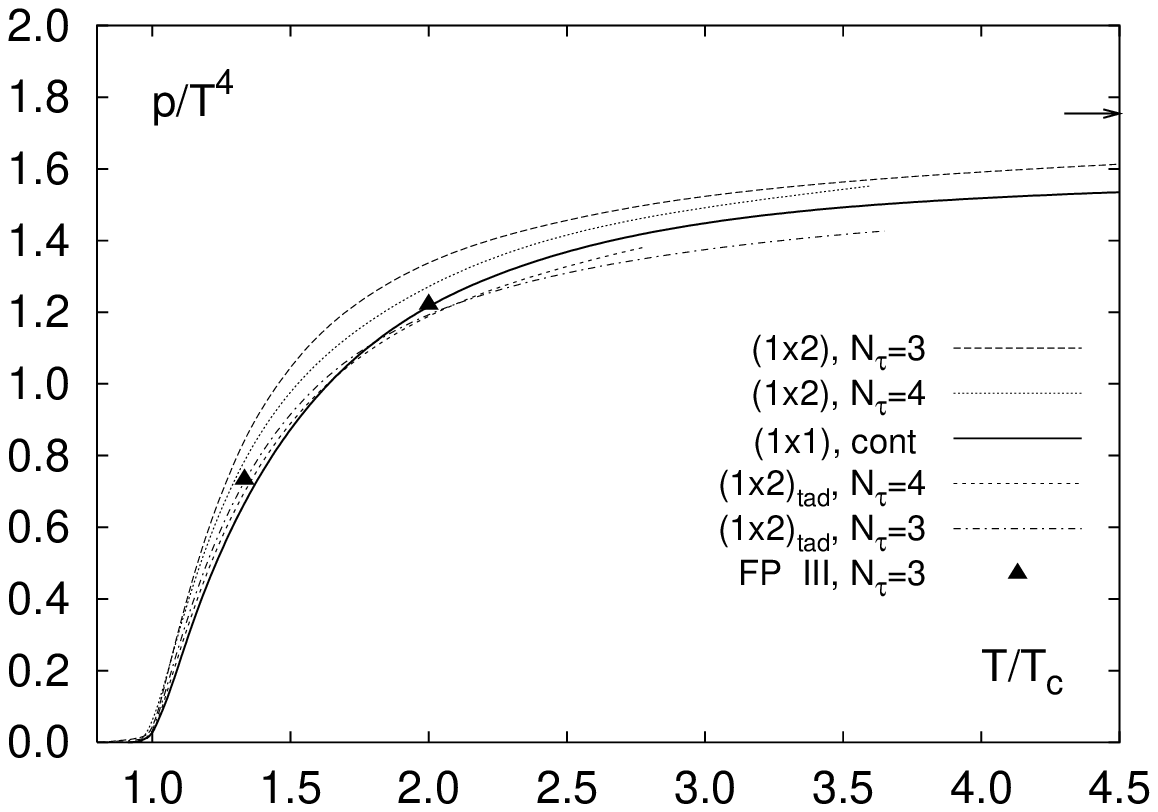,width=80mm}
\end{minipage}
\vskip -0.7truecm
\caption{Pressure of the SU(3) gauge theory calculated with 
the Wilson action and different improved actions on $N_\tau=4$ lattices 
(upper figure). The lower figure shows a comparison of calculations with
tree level and tadpole improved actions on $N_\tau=3$ and 4 lattices.
Also shown there are results from a calculation with a fixed point action 
(triangles).
The arrows indicate the ideal gas result in the continuum limit.  
}
\label{fig:pressure}
\end{figure}

A systematic analysis of the cut-off dependence of bulk thermodynamic 
observables has been performed previously for the Wilson action 
\cite{Boy96}. Calculations on lattices with temporal extent $N_\tau =4$,~6
and 8 could be used there to perform an extrapolation to the continuum limit.
In Fig.~\ref{fig:pressure} we show the results of a calculation of the 
pressure using various improved actions on $N_\tau=4$ lattices.
The strong cut-off dependence for the Wilson action is clearly seen in
the upper plot. While all calculations with improved actions
are close to the continuum extrapolation obtained from the Wilson action
(solid line), the Wilson action result obtained directly on an $N_\tau=4$ 
lattice shows strong deviations (upper curve). In fact, it seems that 
the scattering of the various improved action calculations around the 
continuum extrapolation mainly reflects the 
uncertainties in fixing the temperature scale through calculations of the 
string tension rather than systematic differences in the cut-off dependence.
This is also supported by the lower part of Fig.~\ref{fig:pressure} where
we show a comparison of improved action calculations at different values of the 
cut-off. 
Calculations on $N_\tau=3$ and 4 lattices do not show any
significant cut-off dependence for the tree level
as well as for the tadpole improved actions. 
On the other hand, the spread of results obtained with different actions is 
larger than the cut-off dependence visible for a given improved action.
The results also agree well with a calculation of the
pressure performed with a fixed point action on lattices of 
size $12^3 \times 3$ (triangles) \cite{Pap96}.
It thus seems that systematic errors in the
calculation of the equation of state which result from the lattice
discretization are well under control for the SU(3) gauge theory.

\begin{figure}[htb]
\hskip -0.2truecm  \psfig{
   bbllx=80,bblly=90,bburx=515,bbury=690,
       file=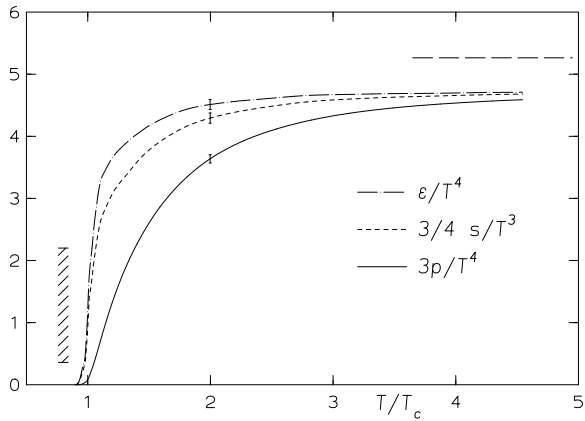, height=80mm, angle=-90}
\caption{Extrapolation to the continuum limit for the energy density,
entropy density and pressure versus $T/T_c$. The dashed horizontal line
shows the ideal gas limit. The hatched vertical band indicates the size of
the discontinuity in $\epsilon/T^4$ (latent heat) at $T_c$.
Typical error bars are shown for all curves.}
\label{fig:continuum}
\end{figure}

In Fig.~\ref{fig:continuum} we show the energy density, pressure and
entropy density obtained from an extrapolation of results obtained
with the standard Wilson action.
We note the rapid rise in all observables at $T_c$ followed by a rather
slow approach to the asymptotic ideal gas behaviour. The latter is in
accordance with the expectation that the high temperature behaviour of
QCD is controlled by a universal function, which only depends on a
{\it running coupling} that varies logarithmically with temperature.
However, although $\epsilon,~s$ and $p$ rapidly come close to the ideal gas 
limit, the (15-20)\% deviations observed at
temperatures as large as $5T_c$ are still too large to be described
by perturbation theory. The perturbative expansion of the
thermodynamic potential is converging quite badly and would require a
running coupling which is significantly smaller than unity \cite{Arn94,Zha95}.
As a consequence, the
perturbative expansion seems to converge only for temperatures much
larger than $T_c$. In particular, the calculation of screening lengths on
the lattice do, however, suggest that the running coupling is larger than
unity even at $5T_c$ \cite{Hel95}.

\subsection{Surface tension and latent heat}

The success of improved actions for the calculation of bulk
thermodynamics even at temperatures close to $T_c$ naturally leads to
the question whether these actions also do lead to an improvement at
$T_c$ itself where a strong cut-off dependence has been observed previously in
the calculation of the latent heat ($\Delta \epsilon$) and the surface
tension ($\sigma_I$) in studies with the Wilson action on lattices with 
temporal extent up to $N_\tau=6$ \cite{Iwa92,Iwa94}.
The region around $T_c$ is, of course, a highly non-perturbative regime.
However, observables like $\Delta\epsilon$ and $\sigma_I$,
which characterize the discontinuities at the first
order deconfinement phase transition in a
$SU(3)$ gauge theory, do depend on properties of the low as well as the
high temperature phase. As the latter is largely controlled by high
momentum modes it may be expected that some improvement does result
even from tree level improved actions.

We have extracted $\sigma_I$ \cite{Bei96b} from the probability
distribution of the absolute value of the Polyakov loop, $P(|L|)$,
following the
analysis presented in Ref.~\cite{Iwa94}. The probability distribution at the
minimum is proportional to
\begin{equation}
P(|L|) \sim \exp\bigl(- \bigl[f_1 V_1 +f_2 V_2 + 2 \sigma_I A\bigr]/T \bigr)
\label{prob}
\end{equation}
where $f_i$ denotes the free energy in the phase $i$, $V_i$ is the
volume occupied by that phase and $A$ denotes the interface area of the
finite system. The distribution functions for Polyakov loop at $T_c$
show a double peak structure. From the depth of the minimum between these
peaks one can extract the surface tension.

\begin{figure}[htb]
\vspace{9pt}
\begin{center}
   \epsfig{
       file=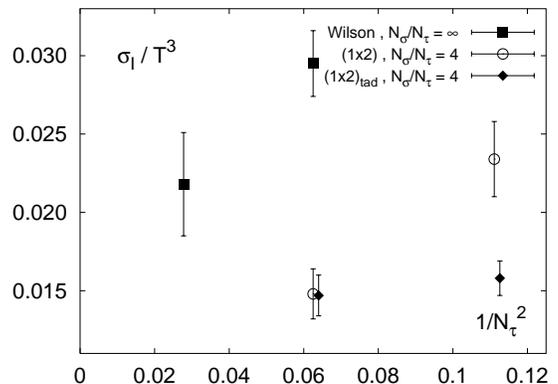,width=80mm}
\end{center}
\vskip -0.7truecm
\caption{Comparison between surface tension calculations performed with 
the Wilson action as well as tree level and tadpole improved (1,2)-actions.  
}
\label{fig:ltadpole}
\end{figure}

In Fig.~\ref{fig:ltadpole} we show results for the surface tension
calculated with the Wilson action as well as tree level and tadpole improved
(1,2)-actions at different values of the cut-off. It is quite remarkable
that the results for the tadpole improved action show practically no
cut-off dependence when comparing calculations on $N_\tau=3$ and 4 lattices. 
Furthermore
it is evident that both improved actions lead to significantly smaller
values for the surface tension at finite values of the cut-off. In fact,
the results for $N_\tau=4$ are in agreement with a quadratic extrapolation
of the Wilson data to the continuum limit.     

In Table~\ref{tab:surface} we give results for $\sigma_I$ on the largest
lattices considered. Clearly the surface tension extracted
from simulations with improved actions on lattices with temporal extent
$N_\tau=4$ are substantially smaller than corresponding results for the
Wilson action. In fact, they are compatible with the $N_\tau=6$ results for
the Wilson action. While the cut-off dependence of the surface tension
calculated on finite lattices seems to be well under control now through
the use of improved actions, it remains to get control over the behaviour in 
the thermodynamic limit ($N_\sigma \rightarrow \infty$). The difference 
between the surface tension calculated on the largest spatial lattice and
the extrapolated value is about 20\%. However, it is not quite clear whether
the ansatz used to perform the extrapolation to the infinite volume is already 
applicable for the currently used spatial lattice sizes. It includes, for
instance, the contribution of translational zero modes, which are 
not yet observed in the present calculations \cite{Iwa94}. Given these 
caveats we find from an extrapolation of the results obtained with the 
tadpole improved actions \cite{Bei96b}
\begin{equation}
{\sigma_I \over T_c^3} = 0.0155 \pm 0.0016
\label{sigmaI}
\end{equation}
\begin{table*}[hbt]
\setlength{\tabcolsep}{0.9pc}
\catcode`?=\active \def?{\kern\digitwidth}
\caption{Surface tension and latent heat for improved actions and
the Wilson action. Results for the Wilson action are based on data from
\cite{Iwa94} using the non-perturbative $\beta$-function calculated 
in \cite{Boy96}. Improved action results are taken from \cite{Bei96b}.
}
\vskip 5pt
\label{tab:surface}
\begin{center}
\begin{tabular}{|l|c|c|c|c|}\hline
action&$V_\sigma$&$N_\tau$&$\sigma_I/T_c^3$&$\Delta\epsilon/T_c^4$
\\ \hline
standard Wilson &$24^2\times 36$&4&0.0300~(16)&2.27~(5) \\
&$36^2\times 48$&6&0.0164~(26)&1.53~(4) \\ \hline
(1,2)-action (tree level) &$32^3$&4&0.0116~(23)&~1.57~(12) \\
(1,2)-action (tadpole) &$32^3$&4&0.0125~(17)&~1.40~(9) \\ \hline
\end{tabular}
\end{center}
\end{table*}

The latent heat is calculated from the discontinuity in
($\epsilon -3p)$. This in turn is obtained from the
discontinuity in the various Wilson loops entering the definition of
the improved actions. In the case of actions, which are defined through 
a set of couplings, which do not depend on the gauge coupling $g^2$, 
the latent
heat is, in fact, simply proportional to the discontinuity in the action 
expectation value at $T_c$. In general one has, however, to take into account
contributions resulting from derivatives of the couplings $c_i(g^2)$
with respect to $g^2$. The latent heat is then given by
\begin{eqnarray}
{\Delta\epsilon \over T_c^4} &=&  {1\over 6}
\biggl({N_\tau \over N_\sigma} \biggr)^3 \biggl( a{{\rm d}\beta \over {\rm
d} a}
\biggr) \biggl(\langle \tilde{S} \rangle_+  - \langle \tilde{S} \rangle_-
\biggr) 
\label{latent}
\end{eqnarray}
with $\tilde{S} \equiv  S - {\rm d} S / {\rm d} \beta$.

The discontinuity in the expectation value of $\tilde{S}$ at $\beta_c$ is 
obtained by calculating these separately in the two coexisting phases 
at $\beta_c$.  In order to extract the latent heat one 
still needs the $\beta$-function entering the definition of $\Delta
\epsilon /T_c^4$ in Eq.~\ref{latent}. The necessary relation $a(\beta)$ has
been obtained from a calculation of $\sqrt{\sigma}a$ (improved actions) or
a determination of $T_ca$ (Wilson action) \cite{Bei96b}.
Results for $\Delta\epsilon/T_c^4$ on $N_\tau = 4$ and 6 lattices 
are summarized in Table~\ref{tab:surface}.

A comparison  with simulations on  even coarser lattices ($N_\tau =3$) 
\cite{Bei96b} shows that the cut-off dependence in the latent heat follows 
a similar pattern as that shown for the surface tension in 
Fig.~\ref{fig:ltadpole}. In particular, the results obtained with the
tadpole improved action on $N_\tau =3$ and 4 lattices agree with each
other within errors. 
We thus may use these results to estimate the latent heat in
the continuum limit. Using Eq.~\ref{tcsigma} and $\sqrt{\sigma}=420~{\rm
Mev}$ we find
\beqn
\Delta \epsilon \simeq 1.5 T_c^4 = 0.23 \sigma^2 = 0.9~{\rm GeV/fm}^3
\label{estimate}
\eqn

\section{Four-flavour QCD with an improved staggered action}

In the fermionic sector of QCD the influence of a finite cut-off on bulk
thermodynamic observables is known to be even larger than in the pure gauge
sector. For instance, in the staggered formulation the energy density of an
ideal fermi gas differs by more than 70\% from the continuum value on a
lattice with temporal extent $N_\tau=4$ and approaches the continuum value
only very slowly with increasing $N_\tau$. This cut-off dependence can
drastically be reduced with an $O(a^2)$ improved staggered action. In a first
attempt to analyze the importance of improvement in the fermion sector we
have performed  calculations with  an improved 
action, $S^I[U] = S^{(1,2)} + \bar{\psi} M \psi$, where a higher order
difference scheme (one-link and three-link terms), is used to improve the
fermionic part \cite{Nai89}. The improved fermion matrix reads
\begin{eqnarray*}
M[U]_{xy}\hskip -0.25cm &=&\hskip -0.2cm \mymatrix{x}{y}  \\
\hskip -2.2truecm {\rm with}\hfill & & \\
A[U]_{xy}\hskip -0.25cm
&=&\hskip -0.1cm \apart{x}{y}{\mu}     \\
B[U]_{xy}\hskip -0.25cm &=&\hskip -0.2cm
\bparta{x}{y}{\mu}  \\
& &\hskip -0.1cm - \bpartb{x}{y}{\mu}
\end{eqnarray*}
where $\eta_{x,\mu}=(-1)^{x_0+..+x_{\mu-1}}$ and $\eta_{x,0}\equiv 1$ denote
the staggered phase factors.
In the gluonic sector we use the tree level improved (1,2)-action.
With this action the overall cut-off distortion of the ideal gas limit
on a
$16^3\times 4$ lattices reduces to about 20\%. We have performed simulations
for two quark masses, $ma =0.05$ and 0.1 \cite{Eng96}. Like in
the pure gauge sector the improvement is visible already close to $T_c$.

\begin{figure}[htb]
\vskip -0.7truecm
\vspace{9pt}
   \epsfig{
       file=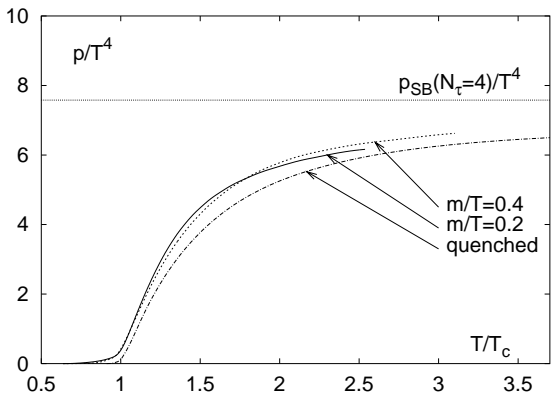,width=80mm}
\vskip -0.7truecm
\caption{Comparison between the pressure of four-flavour QCD on a
$16^3\times 4$ lattice for two values of the quark mass and the pressure of
the pure $SU(3)$ gauge theory. The latter has been re-scaled by the 
appropriate number of degrees of freedom of four-flavour QCD (29/8) and a
factor 1.19 which takes care of the remaining cut-off distortion of the
ideal gas limit resulting from the use of the Naik action.}
\label{fig:pressure4}
\end{figure}

The general structure of the four-flavour equation of state does not
differ much from that of the pure gauge theory. In fact, the temperature
dependence of the pressure is very similar to that of the pure gauge
theory, if we re-scale the latter by an appropriate ratio of the number
of degrees of freedom so that the high temperature limit coincides for
both cases. This is shown in Fig.~\ref{fig:pressure4}.

While the pressure is obtained in complete analogy to the pure gauge
theory through an integration of action differences (Eq.~\ref{preslat}),
the calculation of the energy density now also requires the determination of
differences of the chiral condensates at zero and non-zero temperature
as well as the cut-off dependence of the
two bare couplings, $\beta$ and $m_q$,
\begin{eqnarray}
{\epsilon - 3p \over T^4} &=& - N_\tau^4\biggl[ {{\rm d}\beta \over {\rm d}\ln
a} \biggl(S_0 -S_T \biggr) \nonumber \\
& & + {{\rm d} ma \over {\rm d}\ln a}
\biggl(\langle \bar\chi \chi \rangle_0 - \langle \bar\chi \chi
\rangle_T\biggr)\biggr]
\label{delta4}
\end{eqnarray}
Only in the chiral limit the derivative ${\rm d}\ln ma /{\rm d}a$
vanishes
and $(\epsilon -3 p)$ is again proportional only to the
$\beta$-function, ${\rm d}\beta /{\rm d}\ln a$,
as it is the case in the pure gauge sector.

\begin{figure}[htb]
\vskip -0.7truecm
\vspace{9pt}
   \epsfig{
       file=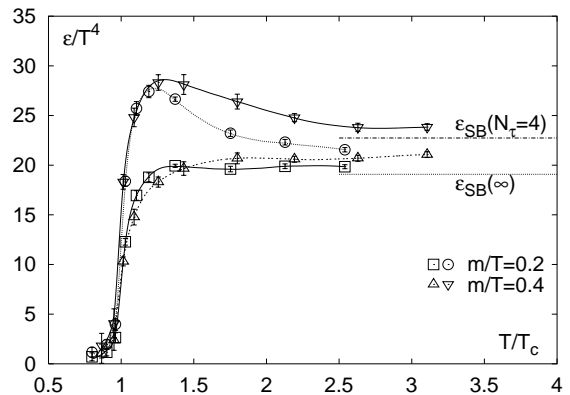,width=80mm}
\vskip -0.7truecm
\caption{Energy density of four-flavour QCD on a $16^3\times 4$ lattice.
The lower set of curves shows an "extrapolation" to the chiral limit
which has been obtained by ignoring the second term in Eq.~10 (see
text).}
\label{fig:energy}
\end{figure}

Results for the energy density are shown in Fig.~\ref{fig:energy}. The
energy density does stay close to the ideal gas limit immediately above
$T_c$. We
do observe an overshooting of the ideal gas limit close to $T_c$ for the
non-zero quark masses con\-sidered by us. This is a feature not seen
before in the pure gauge sector. Whether this will persist for finite
values of the quark mass or is an artifact of our present statistical
accuracy has to be clarified in further more detailed investigations.
The overshooting seems, however, to disappear in an "extrapolation" to
the chiral limit, which we constructed by ignoring the term
proportional to $\langle \bar{\psi} \psi \rangle$ in the definition of the
energy density \cite{Eng96}. The contribution of this term vanishes in
the chiral limit as the derivative
${\rm d}\ln ma /{\rm d}a$ is proportional to the quark mass.

We note that the energy density in the critical region is larger than in the
pure gauge case when expressed in units of $T_c$, {\it i.e.} $\epsilon \sim
10T_c^4$. However, as the critical temperature is substantially smaller than
in the pure gauge theory, the critical energy density again turns out to
be similar, {\it i.e.} about $1$~GeV/fm$^3$.

\section{Improved Fermion Actions}

The calculations with an improved fermion action show that a strong
reduction of the cut-off dependence is possible in the high temperature
phase. Still a further improvement is necessary in order to reduce the
cut-off dependence to only a few percent as it is the case in the pure
gauge sector.
 
Although the Naik action significantly reduces the cut-off dependence 
of thermodynamic quantities calculated with the staggered fermion action
the deviations from the continuum result are still about 20\% on lattices
with temporal extent $N_\tau=4$. Moreover it seems that the flavour
symmetry, which is broken at ${\cal O} (a^2)$ in the staggered fermion 
action, is not
much improved in the Naik formulation \cite{Ber97}. It thus is desirable
to look for more suitable fermion actions. In general, we may discretize 
the free fermion action not only by introducing nearest neighbour couplings
between fermion fields but we may also allow for couplings between sides
further apart,
\begin{eqnarray}
S_F &=& \sum_{x,\mu} \eta_{x,\mu} \bar{\psi}(x) 
 \sum_{j > 0,k,l,m} c_{j,k,l,m} \cdot \nonumber \\ 
& &\cdot \bigl[\psi(x+ja_\mu+ka_\nu
+la_\rho+ma_\sigma)
\nonumber \\
& & 
-\psi(x-ja_\mu-ka_\nu-la_\rho-ma_\sigma) \bigr]~.
\label{naive}
\end{eqnarray}
A specific choice of couplings $c_{j,k,l,m}$ has been given in \cite{Bie97}
for a staggered fixed point action. This does, however, include too many
non-zero couplings for being useful in a numerical simulation.
\begin{table}[t]
\catcode`?=\active \def?{\kern\digitwidth}

\caption{Coefficients for rotationally invariant staggered fermion actions} 
\vskip 5pt
\label{tab:rot}
\begin{center}
\begin{tabular}{|l|l|r|}\hline
$(j,k,l,m)$&\multicolumn{2}{|c|}{$c_{j,k,l,m}$} \\
\hline
~&p4-action&p6-action \\
\hline
\hline
(1,0,0,0) & 0.375     &  0.32       \\
(1,2,0,0) & 0.0208333 &  0.02       \\
(1,2,2,0) &           &  0.0010938  \\
(3,0,0,0) &           &  0.0047917  \\
(1,2,2,2) &           &  0.00125    \\
(3,2,0,0) &           &  0.00125    \\
\hline
\end{tabular}
\end{center}
\end{table}

\begin{figure}[htb]
\vskip -0.7truecm
\vspace{9pt}
   \epsfig{
       file=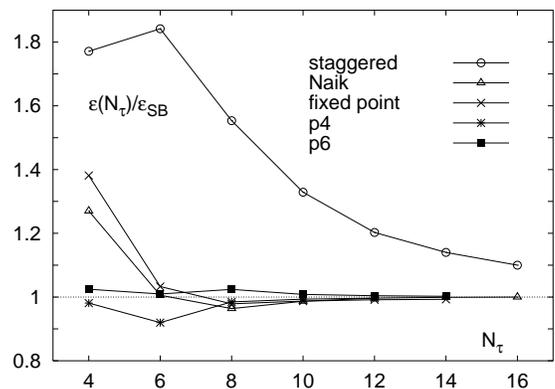,width=80mm}
\vskip -0.7truecm
\caption{Cut-off dependence of the fermion energy density calculated from
several improved actions.}
\label{fig:energy_improved}
\end{figure}

From the analysis of the 
improved gluon actions involving six-link operators we learned that
non-planar loops are, in fact, more efficient in reducing the cut-off
dependence, even when the ${\cal O} (a^2)$ corrections are not eliminated
exactly. One thus may consider to replace the straight three-link term in 
the Naik action by a bended three link term. The coefficient of this term
can be fixed by demanding that the fermion propagator is rotationally 
invariant up to ${\cal O} (p^4)$. Similarly one can construct actions
which include longer paths and are rotationally invariant up to ${\cal O} (p^6)$. 
We call these actions p4 and p6-action, respectively. For instance, in order
to obtain a fermion propagator, which is rotationally invariant up to 
${\cal O} (p^4)$ within the class of staggered fermion actions that contain
non-zero couplings only for one and three-link paths, the coefficients are
constrained by
\begin{eqnarray}
& &c_{1,0,0,0} + ~3 c_{3,0,0,0} + 6 c_{1,2,0,0} = {1 \over 2}\nonumber \\
& &c_{1,0,0,0} + 27 c_{3,0,0,0} + 6 c_{1,2,0,0} = 24 c_{1,2,0,0}~.
\label{threelink}
\end{eqnarray}  
Similar constraints can be obtained for actions that are rotationally
invariant up to ${\cal O} (p^6)$.  A possible set of 
non-zero couplings for these actions is given in Table~3 and a
calculation of the fermion energy density is displayed in 
Fig.~\ref{fig:energy_improved}. We note that with the simple choice of the
p4-action, which includes in addition to the standard one-link term only 
a bended three-link term, the cut-off dependence is reduced to a few percent even 
on $N_\tau=4$ lattices. As the introduction of gauge fields in this action 
will require smeared paths like in the fat link action proposed in \cite{Blu97}
there is the possibility that such actions also improve the 
flavour symmetry of the staggered fermion formulation. This, however, has 
to be examined still in more detail.

\section{Conclusions}

Thermodynamic observables of the $SU(3)$ gauge theory and QCD studied
with improved gauge and fermion actions show a drastic
reduction of the cut-off dependence in the high temperature limit as well
as at $T_c$. The major improvement effect is already obtained with tree
level improved actions. The calculation of the equation of state of the
$SU(3)$ gauge theory seems to be well under control and major
sources for lattice artifacts have been eliminated. At least for finite
temperature calculations on lattices with temporal extent $N_\tau=4$ or
larger systematic effects due to additional tadpole improvement are within
the current statistical accuracy. Tadpole improvement leads to further
reduction of systematic errrors on $N_\tau=3$ lattices.
This opens the possibility to repeat
quantitative studies of various other thermodynamic quantities on rather
coarse lattices which so far could only be investigated on a qualitative
level without reliable extrapolations to the continuum limit.

The calculation of thermodynamic quantities with improved fermion
actions has just started. In this case further work is still needed in
order to be able to select an appropriate action which even in the ideal
gas limit has as little cut-off dependence as the tree level improved
gluon actions.

\vspace{0.5cm}
\noindent
{\bf Acknowledgments:} The work reported here has been performed in  
collaboration with my colleagues at the University of Bielefeld. I would
like to thank them for many discussions and their help in preparing this
talk. In particular I would like to mention B. Beinlich, J. Engels, R. Joswig,
E. Laermann, C. Legeland, M. L\"utgemeier, A. Peikert and B. Petersson.


\begin{thebibliography}{9}
\bibitem{lego} The re-analysis of the string tension data has been 
performed by C. Legeland.
\end{thebibliography}
\end{document}